\documentclass[twocolumn,amssymb,fleqn, twocolumns]{revtex4} 
\usepackage{epsfig,amssymb,amsmath,graphicx,subfigure,hyperref  }  
\usepackage{color}

\usepackage{multirow} 
\usepackage{tabularx}

\begin{document}

\title{Collective dynamics and shattering of disturbed two-dimensional Lennard-Jones crystals}
\author{Zhenwei Yao}
\email[]{zyao@sjtu.edu.cn}
\affiliation{School of Physics and Astronomy, and Institute of Natural Sciences, Shanghai Jiao Tong University, Shanghai 200240, China}
\begin{abstract} 
  Elucidating collective dynamics in crystalline systems is a common scientific
  question in multiple fields. In this work, by combination of high-precision
  numerical approach and analytical normal mode analysis, we systematically investigate the
  dynamical response of two-dimensional Lennard-Jones crystal as a purely
  classical mechanical system under random
  disturbance of varying strength, and reveal rich microscopic dynamics.
  Specifically, we observe highly symmetric velocity field composed of sharply
  divided coherent and disordered regions, and identify the order-disorder dynamical
  transition of the velocity field. Under stronger disturbance, we reveal the
  vacancy-driven shattering of the crystal. This featured disruption mode is
  fundamentally different from the dislocation-unbinding scenario in two-dimensional melting.
  We also examine the healing dynamics associated with vacancies of varying
  size.  The results in this work advance our understanding about the formation
  of collective dynamics and crystal disruption, and may have implications in
  elucidating relevant non-equilibrium behaviors in a host of crystalline
  systems.
\end{abstract}

\maketitle

\section{Introduction}

Understanding dynamical response of crystalline systems to external disturbance
is a common scientific question arising in crystal
stability~\cite{born1940,rehak2012dynamic,morozov_bbok2013}, piezoelectric
effect~\cite{Ikeda,auld}, crystal
melting~\cite{RevModPhys.60.161,nelson2002defects,williams2006dynamical,gasser2010melting},
dynamic
self-assembly~\cite{grzybowski2000dynamic,grinthal2012steering,matsunaga2011controlling,suzuki2016self-assembly},
and various non-equilibrium
phenomena~\cite{vicsek1995novel,helbing2000simulating,smith2010physics,davies2011cancer,marchetti2013hydrodynamics,keber2014topology,sknepnek2015active,yao2016dressed,lowen2021}.
Especially, the disruption of two-dimensional crystals under thermal agitation
(melting) has been intensively studied in the past
decades~\cite{toxvaerd1978melting,barker1981phase,toxvaerd1983size,RevModPhys.60.161,patashinski2010melting,li2020phase,khrapak2020lindemann},
and continuum elasticity theory (KTHNY theory) based on dissociation of
topological defects has been developed to explain two-dimensional crystal
melting~\cite{Kosterlitz1973,young1979melting,nelson1979dislocation,RevModPhys.60.161,nelson2002defects,xuning2016}.
While it has been reported from the perspective of solid mechanics that
instability with respect to a shearing deformation is crucial for the melting
process~\cite{born1940}, further investigation from the perspective of
microscopic dynamics, which has not been fully carried out for the challenge in
direct observation of the phase transition of atomic or molecular materials, may
yield new insights into the fundamental inquiries into the crystal disruption
phenomenon in general~\cite{hwang2019direct}. To this end, the model system of
Lennard-Jones (L-J) crystal, as fabricated by particles interacting under the
L-J potential, provides a suitable
platform~\cite{israelachvili2011intermolecular,yao2014dynamics2,yao2017topological}.
The L-J potential has been extensively employed to simulate interatomic
interactions since 1924~\cite{jones1924determination}.

The goal of this work is to explore the dynamical response of the
two-dimensional L-J crystal system to random disturbance of varying strength,
focusing on the stimulated collective dynamics and the disruption of the crystal.
Specifically, we disturb the system by imposing a random displacement on each
particle. In this work, the L-J lattice is treated as a purely classical
mechanical system. The evolution of the system conforms to deterministic
Hamiltonian dynamics. A suitable
approach to exploring the dynamics of L-J crystals is by high-precision
numerical integration of the equations of motion in combination with theoretical
analysis of normal modes~\cite{rapaport2004art,goldstein2011classical}. This
approach allows us to explore the regime of large disturbance beyond the scope
of harmonic
analysis~\cite{goldstein2011classical,xu2007excess,valles2014molecular,yao2019command}.
The L-J crystal system provides an opportunity to clarify a host of questions
with broad implications, such as: Will the random disturbance lead to coherent
dynamical states? If yes, under which conditions, and which kinds of collective
modes will be excited? How will the crystalline order be disrupted upon strong
disturbance?  Previous studies have shown that geometrically frustrated L-J
crystals can support dislocational and disclinational vacancies due to the steep
energy minimum in the L-J potential curve~\cite{yao2017topological}. It implies
a vacancy-driven disruption scenario in the L-J crystal system that is
distinct from the dislocation-unbinding mechanism in the melting of
two-dimensional crystals~\cite{RevModPhys.60.161,nelson2002defects}.

In this work, we reveal rich microscopic dynamics of two-dimensional L-J crystal
under the disturbance of random orientation and given amplitude. In the perturbation
regime, we find that the velocity field exhibits various featured patterns
composed of sharply divided coherent and disordered regions; the degree of
orderness is characterized by a dynamical order parameter. Remarkably, the
symmetry of the temporally-varying velocity field is well preserved in the dynamical
evolution of the system. Analytical normal mode analysis suggests that the
disordered region originates from the singularity structure in the velocity
field. By increasing the disturbance amplitude, the velocity field experiences a
dynamical transition towards a completely disordered state. Under larger disturbance
amplitude (about $10\%$ of the lattice spacing), we track the dynamical
disruption process, and reveal the vacancy-driven shattering of the
crystal. The finite size effect and stress-free boundary condition are crucial
for shaping such a featured disruption mode. We also examine the healing dynamics
associated with vacancies of varying size. This work reveals rich dynamics in
the L-J model system under random disturbance, and may have implications in
elucidating intriguing non-equilibrium behaviors in a host of crystalline
systems.

\section{Model and method}

The model consists of a collection of point particles confined on the plane
interacting via the L-J potential $V_{LJ}$. In the initial configuration, the
particles constitute a triangular lattice with given boundary shape. The lattice
spacing is set to be the balance distance $r_m$ of the L-J potential. No external
stress is applied on the boundary. Note that
the equilibrium lattice spacing in a finite L-J crystal is slightly smaller than
$r_m$ due to the line tension.  We disturb the system by imposing a random
displacement $\delta \vec{x}$ on each particle. $\delta \vec{x} = b r_m
(\cos\alpha, \sin\alpha)$ in the Cartesian coordinates. The magnitude and
direction of the disturbance is specified by $b r_m$ and $\alpha$, respectively.
$\alpha$ is a uniform random variable in $[0, 2\pi)$. $b$ is a key controlling
parameter, and its value is specified in percentage form. The ensuing evolution
of the system is governed by the Hamiltonian:
\begin{eqnarray}
  H=\sum_{i=1}^{N} \frac{\vec{p}_i^2}{2m_0} + \sum_{i\neq
  j}V_{LJ}(||\vec{x}_i-\vec{x}_j||),
  \label{H}
\end{eqnarray}
where $V_{LJ}(r)=4\epsilon_0[(\sigma_0/r)^{12}
-(\sigma_0/r)^6]$. The balance distance $r_m=2^{1/6}\sigma_0$. We numerically
solve the $2N$ coupled equations of motion derived from eqn (\ref{H}).
Specifically, we employ the Verlet method to construct high-quality
particle trajectories with well conserved total energy, momentum and angular
momentum as well as fixed center of mass for up to a million simulation steps
(see SI for technical details)~\cite{rapaport2004art}. This first-principle
approach allows us to explore the nonlinear regime of large disturbance and
investigate in detail the microscopic dynamics of crystal disruption. In this
work, the units of length, mass, and time are $r_m$, $m_0$, and $\tau_0$,
respectively.  $\tau_0 = r_m\sqrt{m_0/\epsilon_0}$.

\section{Results and discussion}

\begin{figure}[t]  
\centering 
\includegraphics[width=3.7in]{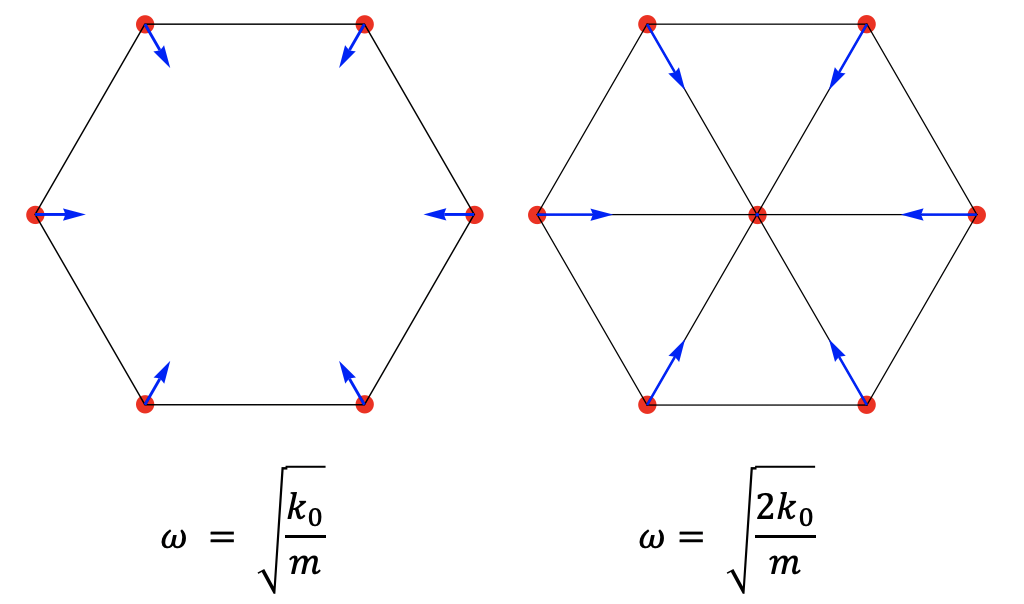}
  \caption{Radial normal modes in the hexagonal 6- and 7-particle configurations
  $H_6$ and $H_7$ that are compatible with the $C_6$ symmetry of the system. The
  black bonds represent springs with stiffness $k_0$. In the $H_6$
  configuration (the right panel), the radial mode has the lowest frequency among all the normal
  modes. In the $H_7$ configuration, the eigenfrequency of the radial mode is
  identical to that of the harmonic oscillator system consisting of two
  spring-connected mass points. }
\label{normal}
\end{figure}

\begin{figure*}[th]  
\centering 
\includegraphics[width=5.8in]{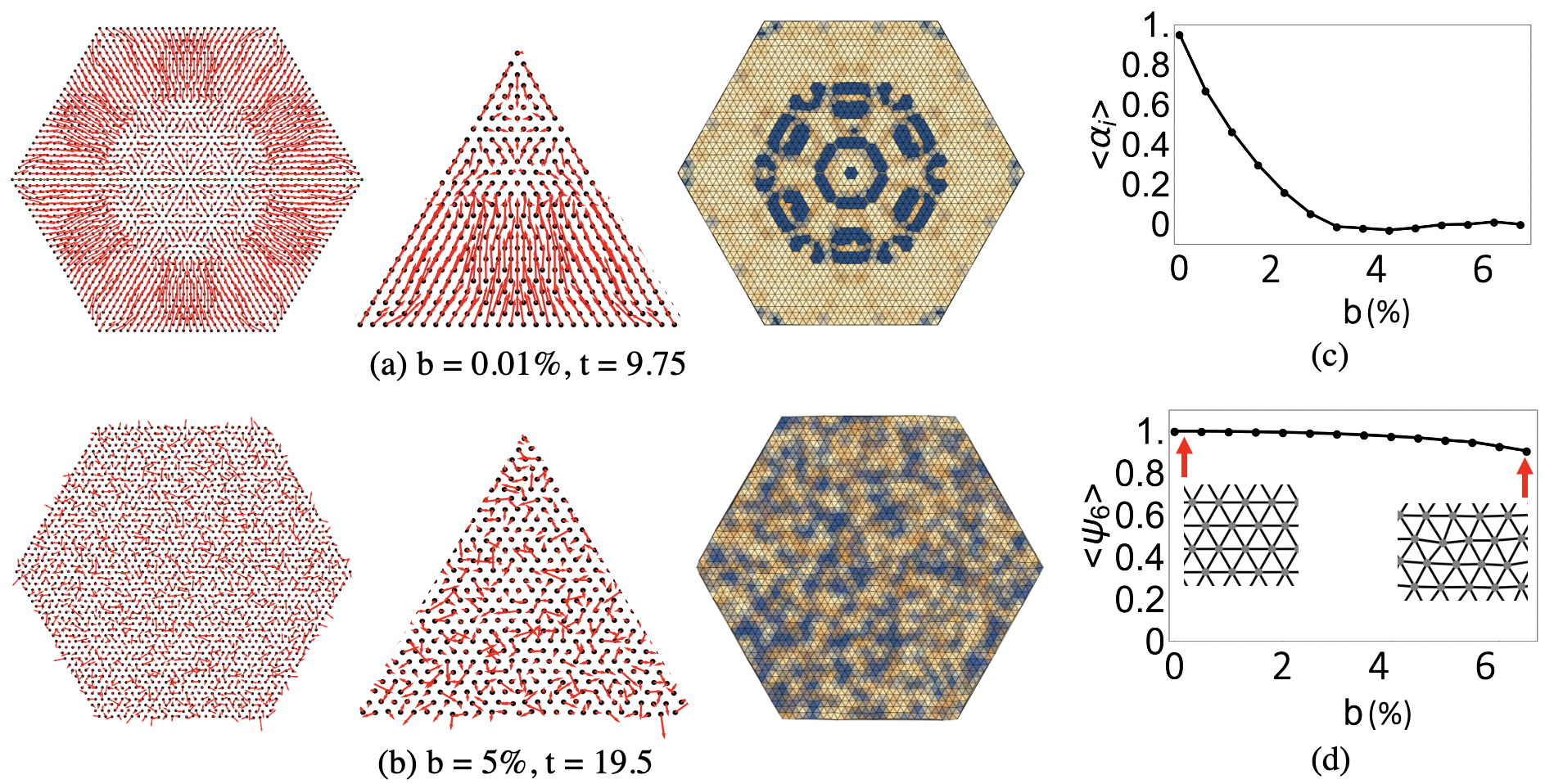}
  \caption{Random disturbance driven dynamical transition in the velocity field
  of a triangular lattice of L-J particles with hexagonal shape and stress-free
  boundary condition.  (a) Coexistence of coherent and disordered regions at the
  disturbance amplitude of $b=0.01\%$. The particle velocity is represented by
  the red arrows. Remarkably, the velocity field maintains
  the $C_6$ symmetry in the evolution of ever-changing pattern. (b)
  Disordered state at $b=5\%$. The panels in the third column are the
  $\alpha-$fields. The dynamical order parameter $\alpha_i$ measures the
  correlation of the velocity of particle $i$ and the average velocity of its
  neighboring particles. The value of $\alpha_i$ is larger in the brighter
  region.  (c) Plot of the spatiotemporally averaged dynamical order parameter
  $\alpha_i$ versus $b$.  (d) Plot of the spatiotemporally averaged $\psi_6$
  versus $b$.  $\psi_6$ characterizes the bond-orientational order of the
  lattice. The spatiotemporal averaging of both $\alpha_i$ and $\psi_6$ is over
  the entire system during the time interval of $100\tau_0$. More instantaneous
  velocity fields are presented in SI.  $N=1951$.
  }
\label{dyn}
\end{figure*}

\subsection{Preliminary analysis of elementary hexagonal configurations}

We first consider the analytically tractable cases of 6- and 7-particle
hexagonal configurations as shown in Fig.~\ref{normal}, and analyze the normal modes in
the perturbation regime, where
the interaction between neighboring particles is approximated by the harmonic
potential. These two kinds of elementary hexagonal configurations are
denoted as $H_6$ and $H_7$. We analytically derive
for all of the normal modes and their frequencies from the eigenvalue
equations~\cite{goldstein2011classical}:
\begin{eqnarray}
  (V_{ij} - T_{ij}\omega_k^2)a_{jk}=0, \label{eom_2}
\end{eqnarray}
where $V_{ij}$ and $T_{ij}$ are the matrices related to the potential and
kinetic energies. $\omega_k$ and $a_{jk}$ are the frequency and eigenvector of
the $k$-mode. The complete results and derivation details are presented in SI.
In all of the normal modes, the total momentum and angular moment are conserved,
and the center of mass is invariant in time.

In the extension of the system from the elementary hexagonal configurations to a
large triangular lattice of hexagonal shape, the analysis of the normal modes in the $H_6$ and $H_7$
configurations inspires a series of inquiries. First of all, among all of the normal
modes, it is found that only the radial modes shown in Fig.~\ref{normal}
possesses the $C_6$ symmetry. Will this symmetry be preserved in the large
system? In general, will the velocity field as a whole respect the
symmetry of the boundary? Furthermore, for the radial mode in the $H_7$ case, it
is noticed that the velocity of the central particle is zero. Does it imply the
appearance of a singularity point in the velocity field in the continuum
limit?  If yes, will the singularity structure in the velocity field manifest
itself as a point or other dynamical morphologies? It is also noticed that the
frequency of the radial mode in the $H_6$ configuration is smaller than that in
the $H_7$ configuration due to the vacancy at the center of the $H_6$ system; in
fact, the radial mode has the lowest frequency among all the normal modes in the
$H_6$ configuration (see SI).  According to Maxwell's rule regarding the
stability of a mechanical structure, removing the central particle in the $H_7$
configuration leads to floppy modes in the resulting $H_6$ configuration, and
thus fundamentally changes the dynamical state of the
system~\cite{maxwell1864calculation,vitelli2012topological}. It is therefore
of interest to examine the vacancy-related dynamics in a large lattice.

\subsection{Formation of collective dynamics and dynamical transition in the
disturbed crystal}

To address the questions proposed in the preceding subsection, we first resort to
high-precision numerical integration of the equations of motion to construct
high-quality particle trajectories with well conserved total energy, momentum
and angular momentum.

In a disturbed triangular lattice of hexagonal shape, we observe that
individual motions of the particles are quickly (in comparison with the
characteristic time $\tau_0$) organized to form coherent dynamical states. A
typical instantaneous velocity field for $b=0.01\%$ (measured in the unit of the
lattice spacing $r_m$) is presented in Fig.~\ref{dyn}(a) for a system of 1951
particles. We see that the entire velocity field exhibits $C_6$ symmetry that is
consistent with the crystal boundary. Remarkably, the velocity field maintains
the $C_6$ symmetry in the evolution of the ever-changing pattern.  We also
investigate the case of rectangular boundaries, and find that the boundary
symmetry also dictates the symmetry of the velocity field. Typical instantaneous
velocity fields of rectangular crystals are presented in SI.

A salient common feature of the velocity field driven by random perturbation is
that the velocity vectors exhibit distinct behaviors in different regions, as
shown in Fig.~\ref{dyn}(a). A zoomed-in plot of a sixth of the entire
$\vec{v}$-field is shown in the second panel. In some region, the motions of the
particles follow the average orientation of the neighboring velocity vectors. In the
remaining region, the orientations of adjacent velocity vectors seem
uncorrelated. To quantify the distinct behaviors of the velocity vectors, we
introduce a dynamical order parameter $\alpha_i$ defined at particle $i$: 
\begin{eqnarray}
  \alpha_i =\frac{ \vec{v}_i\cdot \langle \vec{v}_{j} \rangle} {\| \vec{v}_i \|
  \cdot \| \langle \vec{v}_{j} \rangle \|}. \label{alpha}
\end{eqnarray}
$\langle \vec{v}_{j} \rangle$ is the average velocity of the neighboring
particles of particle $i$; the neighbors of a particle are determined by the
standard Delaunay triangulation~\cite{nelson2002defects}. $\langle \vec{v}_{j} \rangle =
\frac{1}{n_i}\sum_{j=1}^{n_i} \vec{v}_{j}$, where $n_i$ is the coordination
number of particle $i$. When $\vec{v}_i$ follows the orientation of $\langle
\vec{v}_j \rangle$, $\alpha_i=1$, and the cluster of the particles tends to move
coherently. The order parameter $\alpha_i$ thus characterizes the local
alignment of velocity vectors regardless of their magnitude.

In the third panel in Fig.~\ref{dyn}(a), we show the instantaneous
$\alpha$-field with $C_6$ symmetry. In the brighter region, the value of
$\alpha$ is larger. The $\alpha$-field clearly shows that the velocity field is
sharply divided into ordered (bright) and disordered (dark) regions. Here, we
shall note that the degree of order in the velocity field is specified by the
value of the order parameter $\alpha_i$ as defined in eqn (\ref{alpha}); the
system is still in the crystalline state under mild disturbance.  Extensive data
analysis of both hexagonal and rectangular systems shows that the coexistence of
ordered and disordered regions is a common feature of the velocity field in the
perturbation regime. Our proposal of the dynamical order parameter $\alpha_i$
is inspired by the seminal work of active matter physics, where the rule of
specifying particle velocity by the average velocity of surrounding particles
has been employed to create coherent collective
dynamics~\cite{vicsek1995novel}.

The origin of the disordered region in the velocity field is closely related to
the key observation that the boundary particles tend to move perpendicular to
the boundary. Such a boundary dynamical state leads to a nonzero winding number
in the interior velocity vector field~\cite{aminov2000geometry}. Consequently,
singularities are inevitable in the interior velocity field. The disordered
region in the velocity field may be regarded as an area full of singularities in
the continuum limit.  In contrast, by fixing the positions of the boundary
particles, our simulations show that the entire velocity field becomes
disordered everywhere. Note that the disordered region in the velocity field of
the disturbed L-J lattice is distinct from the node structure arising in
standing waves.

With the increase of the disturbance amplitude, we observe the dynamical
transition of the velocity field from the coexistence of ordered and disordered regions to completely disordered state.  In Fig.\ref{dyn}(b), we show that when
the value of $b$ increases to $5\%$ of the lattice spacing, both the velocity
vectors and the $\alpha$-field become disordered. We characterize this dynamical
transition in the plot of $\langle\alpha_i\rangle$ versus $b$ in
Fig.\ref{dyn}(c).  $\langle\alpha_i\rangle$ is the spatiotemporal averaging of
the order parameter $\alpha_i$ over tens of instantaneous velocity fields during
the time interval of $100\tau_0$. From
Fig.\ref{dyn}(c), we see that the velocity field becomes completely disordered
($\langle\alpha_i\rangle=0$) when the amplitude of disturbance exceeds about
$2\%$ of the lattice spacing.  Dynamical transition is also reflected in the
kinetic energy curves: the originally periodic energy curve becomes irregular
at short time scale when dynamical transition occurs (more information about the
energy curves is presented in SI).

Here, we shall emphasize that the crystalline order is still well preserved in
the dynamical transition of the velocity field in preceding discussions. The lattice becomes slightly
wiggly with the increasing disturbance strength. The resulting variation of the
bond-orientational order of the lattice can be quantified by the order
parameter
$\psi_6(\vec{x}_i)$~\cite{Bruinsma1981}:
\begin{eqnarray} 
  \psi_6(\vec{x}_i) = \frac{1}{n_i} \sum_{j=1}^{n_i}e^{i6\theta_j},
\end{eqnarray}
where $\theta_j$ is the angle of the bond connecting the particle $i$ and its
neighbor $j$ with respect to some chosen reference line. $n_i$ is the
coordination number of the particle $i$. From Fig.\ref{dyn}(d), we see the slight
decline of the spatiotemporally averaged $\psi_6$ with the increase of $b$,
indicating that the crystalline order is well preserved up to $b \approx 6\%$.

\begin{figure*}[th]  
\centering 
\includegraphics[width=5.8in]{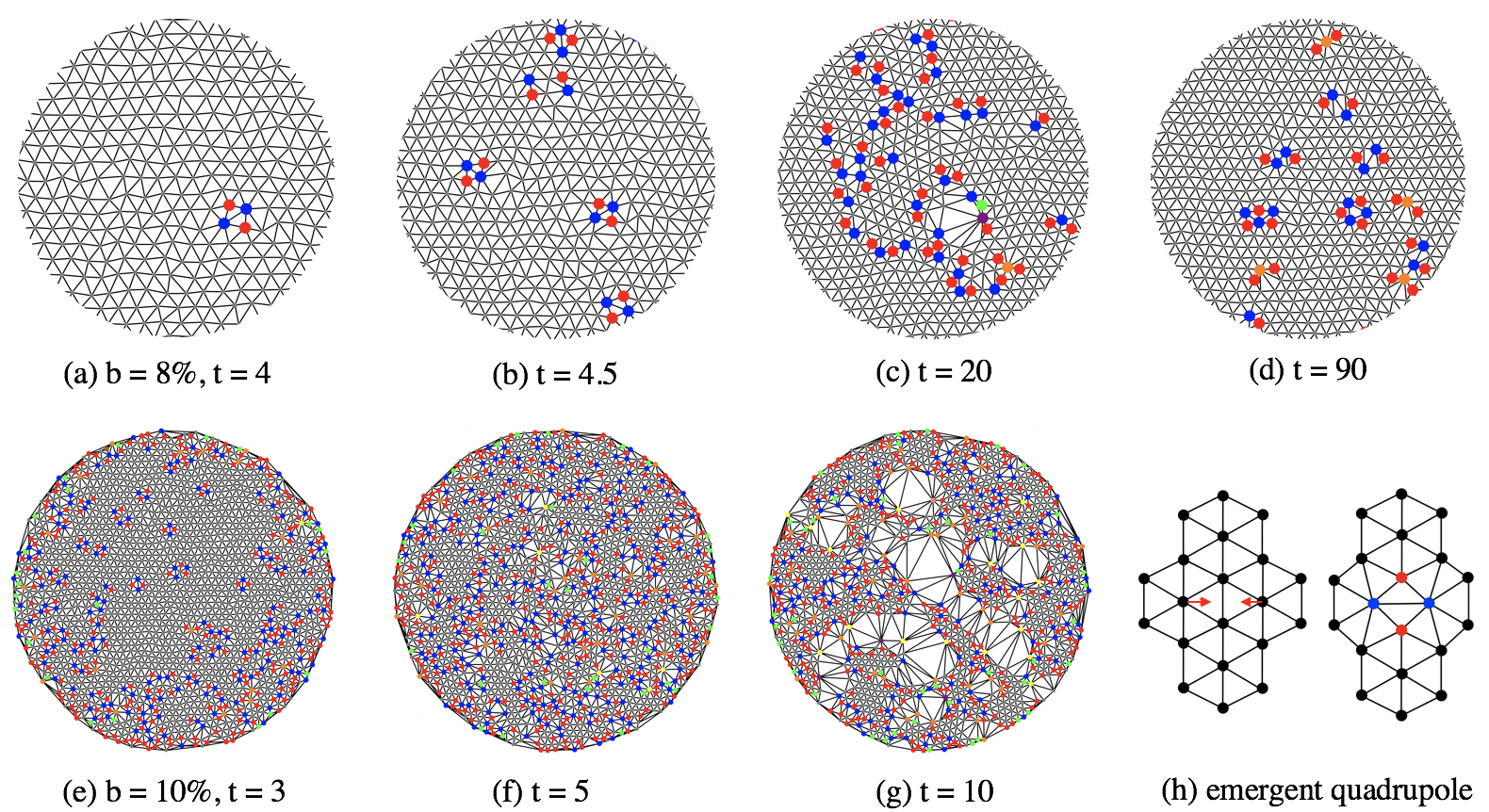}
  \caption{Disruption of the crystalline order in a triangular lattice of L-J
  particles with hexagonal shape and stress-free boundary condition under random
  disturbance. The central circular region of the hexagonal crystals of
  instantaneous particle configurations are shown for clarity. The green, red,
  blue, and orange dots represent four-, five-, seven- and eight-fold
  disclinations, respectively. We observe the sequential emergence of
  quadrupoles, dislocations and void regions (bubbles) in the disturbed lattice.
  Competition of the line tension and the disturbance strength leads to the
  closure of the bubbles at $b=8\%$ (d) and the shattering of the entire crystal
  into pieces of crystallites at $b=10\%$ (g).  (h) Illustration of the
  formation of a quadrupole. $N=1951$.  }
\label{melting}
\end{figure*}

\subsection{Vacancy-driven shattering of the crystal under strong disturbance}

We proceed to explore the regime of stronger disturbance. At $b \approx 7\%$, we
observe the transient existence of a few quadrupoles in the crystal. The
bond-orientational order is still preserved. A schematic plot of a quadrupole is
shown in the right panel in Fig.~\ref{melting}(h). The red and blue dots
represent five- and seven-fold disclinations, which are elementary topological
defects in two-dimensional triangular lattices. An $n$-fold disclination in
triangular lattice is a vertex whose coordination number $n$ is deviated from
six. The coordination number of a particle can be uniquely determined by the
standard Delaunay triangulation~\cite{nelson2002defects}. One may define the
topological charge of an $n$-fold disclination as $q_i=(6-n)\pi/3$.  Topological
charges of the same sign repel and those of opposite signs attract, which
resembles the behavior of electric charges~\cite{nelson2002defects}. A pair of
five- and seven-fold disclinations form a dislocation. And a pair of
dislocations form a quadrupole as shown in Fig.~\ref{melting}(h). While the
emergence of an isolated disclination requires a global reconfiguration of the
particle array, a quadrupole can be excited by a slight local compression of
the particles indicated by the red arrows in Fig.~\ref{melting}(h).
Energetically, it required much less energy to excite a quadrupole than either a
dislocation or a disclination~\cite{Chaikin1995}.

According to the elegant KTHNY theory, the disruption of two-dimensional
crystals under thermal agitation (melting) involves the consecutive
solid-hexatic and hexatic-liquid
transitions~\cite{RevModPhys.60.161,nelson2002defects}. The intermediate hexatic
phase is characterized by the proliferation of dislocations which still
preserves the bond-orientational order; the dislocation-unbinding transition
leads to the isotropic liquid phase.

Here, for our L-J crystal system of finite size under the stress-free boundary
condition, we discover the featured vacancy-driven disruption mode that is
distinct from the dislocation-unbinding mechanism in the melting of
two-dimensional crystals. Typical particle configurations for $b=8\%$ and
$10\%$ are presented in Figs.~\ref{melting}(a)-\ref{melting}(d) and
\ref{melting}(e)-\ref{melting}(g), respectively. Simulations show that the
disruption of the crystalline order is initiated from the center of the system.
For clarity, only the central circular regions of the hexagonal crystals are
shown in Fig.~\ref{melting}. For the case of $b=8\%$, we observe the unbinding
of the quadrupole into a pair of dislocations [see Fig.~\ref{melting}(b)], and
the appearance of void regions (bubbles) extending several lattice spacings [see
Fig.~\ref{melting}(c)]. Line tension emerges once a bubble is formed, and it
tends to inhibit the growing of the bubble. At the disturbance amplitude of
$b=8\%$, the line-tension effect wins, and we observe the closure of the
bubbles, resulting in different kinds of vacancies [see Fig.~\ref{melting}(d);
also see SI]~\cite{yao2017topological}.

By increasing the disturbance amplitude to $b=10\%$, we find that
the line tension fails to suppress the growing of the bubbles, and the entire
crystal is shattered into several pieces of crystallites, as shown in
Fig.~\ref{melting}(e)-\ref{melting}(g). It is noticed that the critical value of
the disturbance amplitude ($b=10\%$) for the mechanical disruption of the
two-dimensional L-J lattice agrees with both the conventional Lindemann melting
criterion for three-dimensional solid~\cite{Lindeman1910} and the modified
Lindemann's criterion for two-dimensional
solid~\cite{zheng1998lindemann,khrapak2020lindemann}. The conventional
Lindemann's criterion states that melting of three-dimensional solid occurs when
the square root of the particle mean-squared displacement from the equilibrium
position reaches a threshold value of about one tenth of the interparticle
distance~\cite{Lindeman1910}. This criterion is generalized to two-dimensional
solid by measuring the displacements of particles in local
coordinate~\cite{zheng1998lindemann,khrapak2020lindemann}. Here, we shall
emphasize that in our work the L-J crystal is two-dimensional, and it is treated
as a purely classical mechanical system whose evolution conforms to
deterministic Hamiltonian dynamics. The Lindemann's criterion is applicable to
the melting of solid as a thermodynamic
system~\cite{zheng1998lindemann,khrapak2020lindemann}.

\begin{figure}[t]  
\centering 
\includegraphics[width=3.5in]{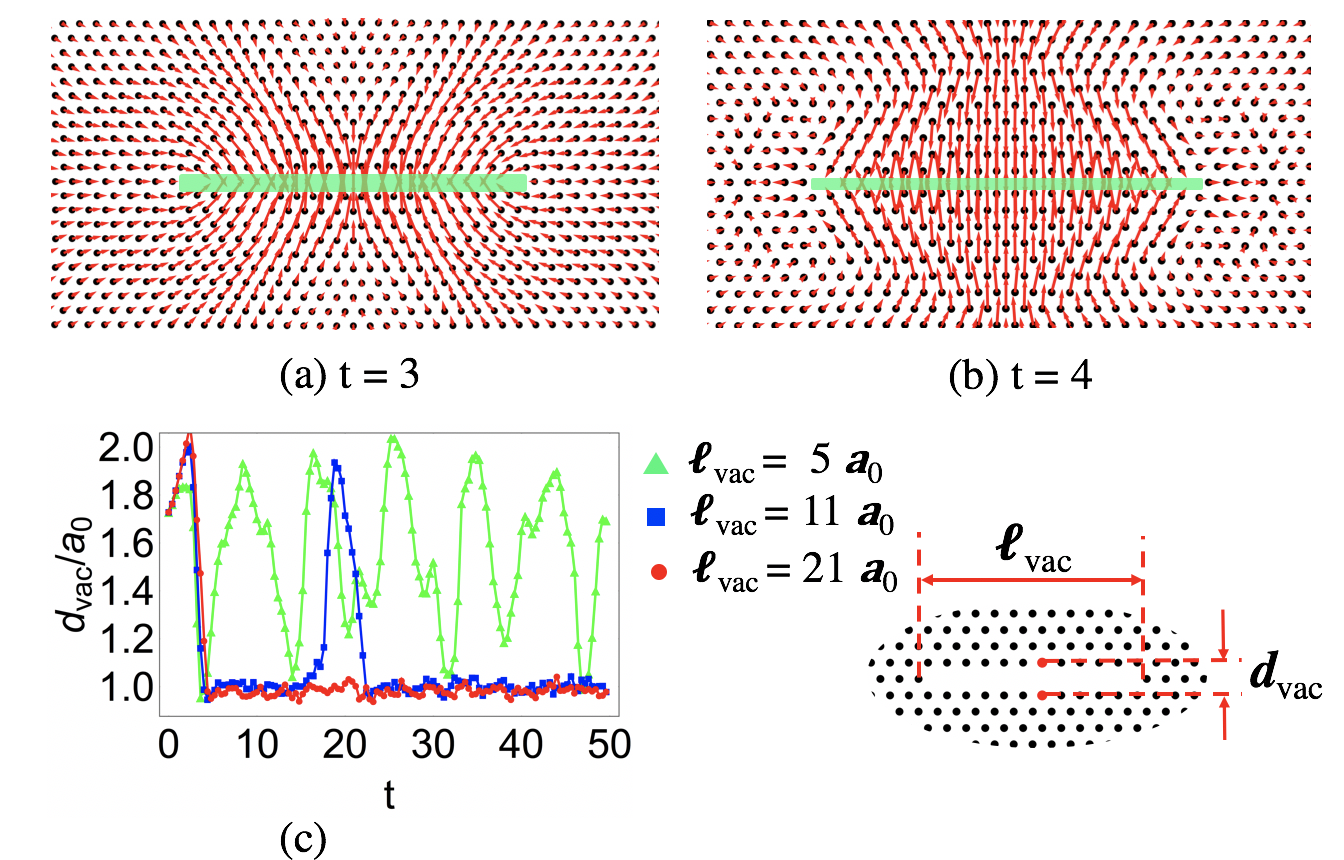}
  \caption{Healing dynamics in a triangular lattice of L-J particles with
  hexagonal shape and stress-free boundary condition. An $n$-point linear
  vacancy is created at the center (indicated by light green bars). For visual
  convenience, only the velocity field near the vacancy is shown. (a) and (b)
  Plots of instantaneous velocity field prior to the closure of the vacancy. (c)
  Plot of the vacancy width $d_{\textrm vac}$ versus time at varying vacancy
  length $\ell_{vac}$. The local particle configuration near the $n$-point
  vacancy is shown in the lower right panel. $b=0$. $N=1951$.
  }
\label{healing_dynamics}
\end{figure}

In contrast, under periodic boundary condition, simulations of L-J systems show
the existence of a metastable state between the solid and liquid
phases~\cite{chen1995melting}, and the featured defect fraction as predicted by
the KTHNY theory~\cite{wierschem2011simulation}. Furthermore, for the case of
a hexagonal L-J crystal with fixed boundary, our simulations show that the
constraint of fixed area suppresses the development of bubble structures, and
the disruption process resembles the melting scenario of two-dimensional
crystals according to the KTHNY theory (more information is presented in SI).
To conclude, the disruption of the two-dimensional L-J crystal occurs in the
form of the vacancy-driven shattering under the stress-free boundary condition,
which is in contrast to the dislocation-unbinding mechanism (the scenario of
the KTHNY theory) at constant density as implemented by either the periodic
boundary condition or the fixed boundary condition.

Fluctuation of the bubble structure in its size inspires us to ask how a void
region is healed under line tension. To address this question, we create an
$n$-point linear vacancy at the center of a hexagonal L-J crystal simply by removing
$n$ particles. Typical instantaneous velocity fields prior to the closure of
the $n$-point vacancy are shown in Figs.~\ref{healing_dynamics}(a) and \ref{healing_dynamics}(b). The
particles near the vacancy move coherently to shrink the width $d_{\textrm {vac}}$ of the
linear vacancy as indicated by the light green bar. Here, the introduction of the linear
vacancy breaks the $C_6$ symmetry of the system. The symmetry of the resulting
velocity field changes accordingly. We track the temporally-varying vacancy
width $d_{\textrm {vac}}$ at varying vacancy length $\ell_{\textrm {vac}}$ (without
imposing any disturbance), and the results are presented in
Fig.~\ref{healing_dynamics}(c). It turns out that the short- and long-vacancies
exhibit distinct dynamical behaviors.  The short vacancy ($\ell_{\textrm {vac}}=5 a_0$)
alternately opens and closes.  In contrast, the long vacancy ($\ell_{\textrm {vac}}=21
a_0$) is quickly healed, leaving out a pair of dislocations at its two ends with
opposite Burgers vectors. Note that our previous molecular dynamics simulations
have shown that infinitely large two-dimensional L-J crystal exhibits similar
behaviors~\cite{yao2014dynamics2}. For the intermediate case of $\ell_{\textrm {vac}}=11
a_0$, our numerical solution captures the transient open state of the vacancy
for a duration of $\Delta t \approx 11$ [see the blue curve in
Fig.~\ref{healing_dynamics}(c)]. By imposing perturbation, the
$d_{vac}(t)$ curves at varying vacancy size for $b=1\%$ are almost identical
with those in Fig.~\ref{healing_dynamics}(c).

\section{Conclusion}
  
In summary, in this work we have explored the random disturbance driven
microscopic dynamics in the L-J crystal system. In the perturbation regime, we
observed the symmetry-preserved velocity field consisting of sharply divided
coherent and disordered regions. Under stronger disturbance, we identified the
dynamical transition, and revealed the featured vacancy-driven disruption of
crystal in the form of shattering. We also discussed the healing dynamics
associated with vacancies of varying size. An important observation in our study
is the crucial role of the boundary condition for shaping the collective
dynamics and the disruption mode.  Specifically, under the stress-free boundary
condition, the boundary particles tend to move perpendicular to the boundary,
which is responsible for both the preservation of the symmetry and the formation
and evolution of the singularity structure in the velocity field. The
stress-free boundary condition also allows the system to develop the bubble
structure, which is closely related to the shattering of the crystal. These
results advance our understanding about collective dynamics and crystal
disruption, and may have implications in elucidating intriguing dynamical
non-equilibrium behaviors in a host of crystalline systems.\\

\noindent \textbf{Acknowledgements}

\noindent This work was supported by the National Natural Science Foundation of
China (Grants No. BC4190050).\\



\bibliographystyle{spphys}       


\end{document}